%% file: DFARXIVV3.tex
\documentclass[journal, final, 10pt, twocolumn]{IEEEtran} 
\usepackage{graphicx,epsf,float}
\usepackage{amsmath,amsfonts,amstext,amssymb,amsbsy,amsopn,amsthm,bm}
\usepackage{ifdraft,subcaption,acronym,color,units,psfrag}
\usepackage[sort]{cite}

\usepackage{balance}
\usepackage{booktabs}
\usepackage{tikz}
\usepackage{pgfplots}
\pgfplotsset{compat=newest}
\usepackage{steinmetz}

\usetikzlibrary{arrows, decorations.markings}
\usepackage{graphicx}

\usepgfplotslibrary{groupplots}

\acrodef{DF}{deep filter}
\acrodef{DOA}{direction-of-arrival}
\acrodef{STFT}{short-time Fourier transform}
\acrodef{PDF}{probability density function}
\acrodef{ULA}{uniform linear array}
\acrodef{SDNR}[segDNR]{segmental diffuse-to-noise ratio}
\acrodef{LSD}{log spectral difference}
\acrodef{SDSR}[segDSR]{segmental diffuse-to-signal ratio}
\acrodef{LCMV}{linearly constrained minimum variance}
\acrodef{MMSE}{minimum mean square error}
\acrodef{WNG}{white noise gain}

\definecolor{myblue}{rgb}{0.2,0.2,0.9}
\definecolor{orange}{rgb}{1,0.27,0}

\usepackage{booktabs}
\usepackage{graphicx}
\usepackage{color}
\usepackage{calc}

 \def\baselinestretch{0.985}

\begin{document}
\title{\huge \bf Deep Filtering: Signal Extraction and Reconstruction Using Complex Time-Frequency Filters}

\author{Wolfgang Mack, and Emanu\"{e}l A.~P.~Habets,~\IEEEmembership{Senior~Member,~IEEE} \thanks{Wolfgang Mack and Emanu\"{e}l Habets are with the International Audio Laboratories Erlangen (a joint institution of the Friedrich-Alexander-University Erlangen-N\"urnberg (FAU) and Fraunhofer IIS), Germany. E-mail: \{wolfgang.mack, emanuel.habets\}@audiolabs-erlangen.de}}

\maketitle

\begin{abstract}
Signal extraction from a single-channel mixture with additional undesired signals is most commonly performed using time-frequency (TF) masks. Typically, the mask is estimated with a deep neural network (DNN), and element-wise applied to the complex mixture short-time Fourier transform (STFT) representation to perform the extraction. Ideal mask magnitudes are zero for solely undesired signals in a TF bin and undefined for total destructive interference. Usually, masks have an upper bound to provide well-defined DNN outputs at the cost of limited extraction capabilities. We propose to estimate with a DNN a complex TF filter for each mixture TF bin which maps an STFT area in the respective mixture to the desired TF bin to address destructive interference in mixture TF bins. The DNN is optimized by minimizing the error between the extracted and the ground-truth desired signal allowing to learn the TF filters without having to specify ground-truth TF filters. We compare our approach with complex and real-valued TF masks by separating speech from a variety of different sound and noise classes from the Google AudioSet corpus. We also process the mixture STFT with notch-filters and zero whole time-frames, to simulate packet-loss during transmission, to demonstrate the reconstruction capabilities of our approach. The proposed method outperformed the baselines, especially when notch-filters and time-frame zeroing were applied.
\end{abstract}

\vspace{-.6em}

\section{Introduction}
Real-world sound signals often consist of a mixture of desired sounds and additional interfering sources like traffic noise, music, or babble speech. Further degradations of the desired signals in the mixture can be caused by preprocessing or specific room geometries, which cause notch-filters, or by packet-loss during transmission. Extracting and reconstructing desired signals from such a mixture is highly desirable in applications like speech enhancement, source separation, or packet-loss concealment. In this paper, we propose a single-channel approach, which extracts a desired signal from a mixture of desired and undesired signals, that consist of interfering sounds and also model additional degradations like notch-filters or packet-loss. Hence, our approach jointly addresses interference reduction and signal reconstruction.

Typically, signal extraction is performed in the short-time Fourier transform (STFT) domain, where the real and imaginary part (e.g., \cite{Tan2019}), or the spectral magnitude (e.g., \cite{Han2015}) of the desired signals are estimated. Other approaches estimate time-frequency (TF) masks, which are consequently element-wise applied to the complex mixture STFT to perform the extraction.
Estimating TF masks is usually preferred over directly estimating spectral magnitudes due to performance reasons \cite{Wang2014}. Typically, TF masks are estimated from a mixture representation by a deep neural network (DNN)  (e.g., \cite{Williamson2016,Williamson2017,Hershey2016,Chen2017,Isik2016,Wang2014, Yu2017,Luo2017,wang2019deep,8664086}) where the output layer often directly yields the STFT mask. Two common approaches exist to train such DNNs. First, a ground-truth mask is defined, and the DNN learns the mixture to mask mapping by minimizing an error function between the ground-truth and estimated masks (e.g., \cite{Williamson2016, Hershey2016}).  In the second approach, the DNN learns the mapping by directly minimizing an error function between the estimated and the desired signal (e.g., \cite{Yu2017,Kolbaek2017,Mack2018}). Erdogan et al. \cite{Erdogan2018} showed that direct optimization is equal to mask optimization weighted with the squared mixture magnitude. Consequently,  the impact of high energy TF bins on the loss is increased, and the impact of low energy decreased. Furthermore, no ground-truth mask has to be defined as it is implicitly given when specifying the desired signal.  

For different extraction tasks, different types of TF masks have been proposed. Given a mixture in STFT domain where the signal in each TF bin either belongs solely to the desired or the undesired signal, extraction can be performed using binary masks \cite{Wang2005a} (e.g.,  \cite{Hershey2016, Isik2016}). Given a mixture in STFT domain where several sources are active in the same TF bin, ratio masks (RMs) \cite{Hummersone20142} or complex ratio masks (CRMs) \cite{Mayer2017} can be applied. Both assign a gain to each mixture TF bin to estimate the desired spectrum. The real-valued gains of RMs perform TF bin-wise magnitude correction from the mixture to the desired spectrum. The estimated phase is, in this case, equal to the mixture phase. CRMs apply a complex instead of a real gain and additionally perform phase correction. Speaker separation, dereverberation, and denoising have been achieved using RM (e.g., \cite{Kolbaek2017, Chen2017,Mack2018, Yu2017,Weninger2015}) and CRM (e.g.\cite{Williamson2016, Williamson2017}). Ideally, the magnitude of RMs and CRMs is zero if only undesired signals are active in a TF bin and much larger than one if the desired and undesired signals overlap destructively in a certain TF bin. Outputs approaching infinity cannot be estimated well with a DNN. For obtaining well-defined DNN outputs, it is possible to estimate a compressed mask (e.g., \cite{Williamson2017}) with a DNN and perform extraction after decompression to obtain mask values with high magnitudes. Weak noise on the DNN output, however, can lead to a huge change in the estimated masks resulting in big errors. Furthermore,  when the desired and undesired signals in a TF bin add up to zero, also a compressed mask cannot reconstruct the respective magnitude from zero by multiplication. Often, the case of destructive interference is ignored (e.g., \cite{ Xu2017,Chen2017,Mack2018}), and mask values bounded to one are estimated because higher values also come with the risk of noise amplification. Besides masks, also complex-valued TF filters (e.g.,  \cite{Benesty2011}) have been applied for the purpose of signal extraction. Current TF filter approaches usually incorporate a statistics estimation step (e.g., \cite{Benesty2011,Benesty2011a,Fischer2016, Doerte2018}), which can be crucial given a large variety of unknown interference signals with fast-changing statistics as present in real-world scenarios.

In this paper, we propose to use a DNN to estimate a complex-valued TF filter for each TF bin in the STFT domain to address extraction also for highly non-stationary signals with unknown statistics. The filter is element-wise applied to a defined area in the respective mixture STFT. The result is summed up to obtain an estimate of the desired signal in the respective TF bin. The individual complex filter values are bounded in magnitude to provide well-defined DNN outputs. Each estimated TF bin is a complex weighted sum of a TF bin area in the complex mixture. This allows addressing the case of destructive interference in a single TF bin without the noise-sensitivity of mask compression. It also allows reconstructing a TF bin which is zero by taking into account neighboring TF bins with non-zero magnitudes. The combination of DNNs and TF filters mitigates both the shortcomings of TF masks and of existing TF filter approaches.

The paper is structured as follows. In Section~\ref{sec:tfmasks}, we present the signal extraction process with TF masks and subsequently, in Section~\ref{sec:filter}, we describe our proposed method. Section~\ref{sec:datasets} contains the data sets we used and Section~\ref{sec:peval} the results of the experiments to verify our theoretical considerations.
\section{STFT Mask based Extraction}
\label{sec:tfmasks}
In this section, we review the extraction process with TF masks and provide implementation details of the masks we used as baselines in the performance evaluation.
\subsection{Objective}
We define the complex single-channel  spectrum of the mixture as $X(n,k)$, of the desired signal  as $X_{\text{d}}(n,k)$, and of the undesired signal as $X_{\text{u}}(n,k)$ in STFT domain where $n$ is the time-frame and $k$ is the frequency index. We consider the mixture $X(n,k)$ to be a superposition
\begin{equation}
\label{Equ:addition}
X(n,k) = X_{\text{u}}(n,k) +X_{\text{d}}(n,k).
\end{equation} 
The objective of mask-based extraction is to obtain an  estimate  of $X_{\text{d}}(n,k)$ by applying a mask to $X(n,k)$, i.e., 
\begin{equation}
\label{Equ:applmask}
 \widehat{X}_{\text{d}}(n,k) =  \widehat{M}(n,k) \cdot X(n,k) ,
\end{equation} 
 where  $\widehat{X}_{\text{d}}(n,k)$ is the estimated desired signal and $\widehat{M}(n,k)$ the estimated TF mask.
 
Usually TF masks are estimated with a DNN which is either optimized to estimate a predefined ground-truth TF mask $M(n,k)$ for all $N\cdot K$ TF bins, where $N$ is the total number of time-frames and $K$ the number of frequency bins per time-frame
\begin{equation}
\label{Equ:maskloss}
 J_{\text{M}} = \frac{1}{N\cdot K} \sum_{k = 1}^K \sum_{n = 1}^N |  M
 (n,k) -  \widehat{M
}(n,k)|^2,\end{equation} 
or to reduce the reconstruction error between $X_{\text{d}}(n,k)$ and  $\widehat{X}_{\text{d}}(n,k)$
 \begin{equation}
 \label{Equ:comploss}
 J_\text{R} = \frac{1}{N\cdot K} \sum_{k = 1}^K \sum_{n = 1}^N | ( X_{\text{d}}(n,k) -  \widehat{X}_{\text{d}}(n,k))^2|, 
 \end{equation} 
 or to reduce the magnitude reconstruction
  \begin{equation}
 \label{Equ:ratioloss}
 J_\text{MR} = \frac{1}{N\cdot K}\sum_{k = 1}^K \sum_{n = 1}^N  ( |X_{\text{d}}(n,k)| - | \widehat{X
}_{\text{d}}(n,k)|)^2.
 \end{equation} 
For destructive interference in (1), when
\begin{equation}
\label{Equ:triangle}
|X_\text{d}(n,k) + X_\text{u}(n,k)| \leq |X_\text{d}(n,k)|,
\end{equation}
then the ideal mask magnitude $|M(n,k)|=|\frac{X_{\text{d}(n,k)}}{X_{\text{d}(n,k)}+X_{\text{u}(n,k)}}|$ is  $\in [1, \infty[$. Hence, the global optimum cannot always be reached if $|\widehat{M}(n,k)|$ is bounded.
\subsection{Implementation}
\label{subsec:maskimpl}
For mask estimation, we use a DNN with three bidirectional long short-term memory (BLSTM)  layers \cite{Hochreiter1997} with $1200$ neurons per layer and a feed-forward output layer. We trained DNNs with a linear and a tanh output activation to incorporate masks with and without bounded values in our experiments. The output $O$ is of dimension $(N,K,2)$ representing a real and imaginary output per TF bin. 

For mask estimation, we designed the model to have the same number of trainable parameters for the RM and CRM approaches. We used a real-valued DNN with the stacked real and imaginary part of $X$ as input and two outputs, defined as $O_\text{r}$ and $O_\text{i}$, per TF bin. These can be interpreted as real and imaginary mask components. For RM estimation, we computed $\widehat{M}(n,k)=\sqrt{O_\text{r}(n,k)^2 +O_\text{i}(n,k)^2}$ . 
We trained DNNs to estimate RMs optimized with (\ref{Equ:ratioloss}) and CRMs optimized with (\ref{Equ:comploss}). We computed the complex multiplication of $X(n,k)$ and $\widehat{M}(n,k)$ in (\ref{Equ:applmask}) for the CRMs by
\begin{align}
\label{Equ:commulreal}
&\begin{aligned}
Re\{\widehat{X}_{\text{d}}\} = Re\{\widehat{M}\} \cdot Re\{X\} - Im\{\widehat{M}\} \cdot Im\{X\} ,
\end{aligned}\\ 
&\begin{aligned}
\label{Equ:commulim}
Im\{\widehat{X}_{\text{d}}\} = Im\{\widehat{M}\} \cdot Re\{X\} + Re\{\widehat{M}\} \cdot Im\{X\}.\end{aligned}
\end{align}
Note that $(n,k)$ is omitted for brevity. We trained 100 epochs, used the Adam \cite{Kingma2015} optimizer, a dropout \cite{Srivastava2014} of 0.4 in the BLSTMs, a batch size of 64, a learning rate of 1e-4.
\section{Proposed STFT Filter based Extraction}
\label{sec:filter}
\begin{figure}
\def\svgwidth{\columnwidth }
\input{./FiltervsMask.pdf_tex}
\vspace{-2em}
\caption{Scheme of input and output STFT. The orange cells represent the considered regions of the input STFT to be mapped to the one orange grid in the respective output STFT. In a), there is a one-to-one mapping via a TF mask, and in b), there is a many-to-one mapping via a TF filter. }
\vspace{-1.2em}
\label{fig:maskvsfilter}
\end{figure}
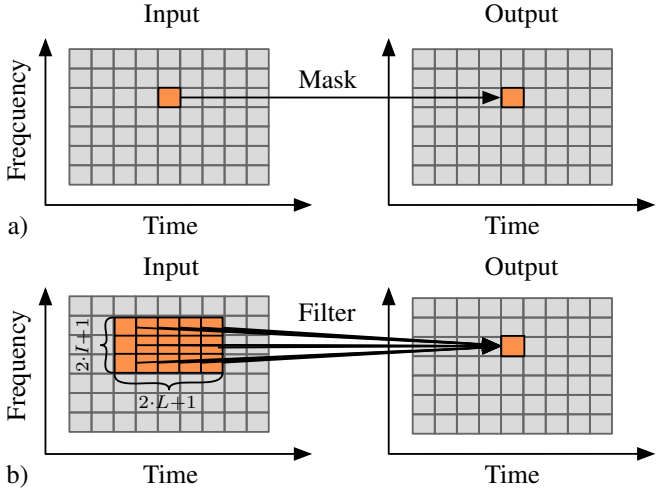
In this section we show how to estimate $X_\text{d}$ using an STFT domain filter instead of TF masks as depicted in Figure \ref{fig:maskvsfilter}. We refer to this filter as a \ac{DF}.
\subsection{Objective}
We propose to obtain $\widehat{X}_{\text{d}}$ from $X$ by applying a complex filter
\begin{equation}
\label{Equ:filterreconstruction}
\widehat{X}_{\text{d}}(n,k) = \sum_{i = -I}^{I} \sum_{l = -L}^{L} H^*_{n,k}(l+L,i+I)\cdot X(n-l,k-i) , 
\end{equation}
where $2\cdot L + 1$ is the filter dimension in time-frame direction and $2\cdot I +1$ in frequency direction and $H^*_{n,k}$ is the complex conjugated 2D filter of TF bin $(n,k)$. Note that, without loss of generality, we used in (\ref{Equ:filterreconstruction}) a rectangular filter centered around $(n,k)$ only for reasons of presentation simplicity. The filter values are bound in magnitude to provide well-defined DNN outputs
\begin{equation}
|H^*_{n,k}(l+L, i+I)| \leqslant b \text{ }\forall l,i \in \mathbb{N}: l,i\in [-L,L],[-I, I], 
\end{equation}
where $b \in \mathbb{R}$ is the bound which depends on the DNN output activation. The DNN is optimized according to (\ref{Equ:comploss}) which allows training without having to define ground-truth filters (GTFs) and to directly optimize the reconstruction mean squared error (MSE). The decision for GTFs is crucial because there are usually infinitely many combinations of different filter values that lead to the same extraction result. If a GTF was selected randomly for a TF bin from the set of infinitely many GTFs, training would fail because there would not be consistency between the selected filters. We can interpret this situation as a partially observable process for the GTF designer and fully observable for the DNN. From the input data properties, the DNN can decide exactly which filter to take without ambiguities. The GTF designer has an infinitely large set of possible GTFs but cannot interpret the input data to decide which GTF to take so that the current DNN update is consistent w.r.t. previous updates. By training with (\ref{Equ:comploss}), we avoid the problem of GTF selection.
\subsection{Implementation}
We used the same DNN as proposed in Section~\ref{subsec:maskimpl} with a tanh output activation to define well-defined DNN outputs and changed the output shape to $(N, K, 2, 2\cdot L + 1,2\cdot I +1)$, where the last $2$ entries are the filter dimensions. The complex multiplication in (\ref{Equ:filterreconstruction}) was performed as shown in (\ref{Equ:commulreal}) and (\ref{Equ:commulim}). We experimented with different filters with $L,  I \in \{0,1,2\}$ resulting in filter dimensions of $(2 \cdot L+1,2 \cdot I+1)$. The filter shapes of the respective models are represented as subindices, $\text{DF}_{2 \cdot L+1 \times 2\cdot I+1 }$. The maximum of $|H_{n,k}(l, i)|$ is phase-dependent $\in [1,\sqrt{2}]$ for the employed tanh activation. As all $|H_{n,k}(l, i)|$ can be at least $1$, a DNN can theoretically optimize (\ref{Equ:comploss}) to its global optimum zero, if
\begin{equation}
\label{Equ:recbound}
c \cdot\sum_{i=-I}^{I} \sum_{l=-L}^{L} |X(n-l, k - i)| \geqslant | X_{\text{d}}(n,k)|,
\end{equation}
where $c \in \mathbb{R^+}$ is the maximal magnitude all filter values can reach. Hence, to address destructive interference, the summation of all mixture magnitudes considered by a filter weighted with $c$ must be at least equal to the desired TF bin magnitude. In our experiments, we used $c=1$.
\section{Data Sets}
\label{sec:datasets}
We used AudioSet  \cite{gemmeke2017} (without the speech samples), containing a large variety of highly non-stationary sound samples,  as interference and LIBRI \cite{Libri2015} as desired speech data corpora. 
All data was downsampled to 8~kHz sampling frequency and had a duration of 5~s. For the  STFT we set the hop size to 10~ms, the frame length to 32~ms, and used the Hann window. Consequently, in our tests $ K = 129$ and $N = 501$ and the temporal context per filter is $(2\cdot L)\cdot 10~\text{ms} +32~\text{ms}$ and the frequency context is $(2\cdot I +1)\cdot 62.5~\text{kHz}$.

For training, validation, and test data generation, we added interference from AudioSet with a segmental signal-to-noise-ratio (SNR) $\in [0,6]$~dB (different samples for training, validation, and test), white noise with an SNR $\in [20,30]$~dB, notch-filtering and, to simulate packet-loss, random time-frame zeroing (T-kill) as degradations in the mentioned order. Each degradation was applied to a sample with a probability of $50$ percent. For notch-filtering, we randomly selected a center frequency with a quality factor $\in [10,40]$. When T-kill was applied, every time-frame was zeroed with a probability of 0.1. We generated 100k training, 5k validation, and 50k test samples. The test samples were divided into four subsets, Test~0, Test~1, Test~2, and Test~3. Test~0 contains clean speech ($\approx$~3k~samples). In Test~1, speech was degraded by interference from AudioSet and white noise ($\approx$~6k~samples). In Test~2, speech was degraded by notch-filtering, T-kill, and white noise ($\approx$~6k~samples). In Test~3, speech was degraded by interference from AudioSet, notch-filtering, T-kill, and white noise simultaneously ($\approx$~6k~samples). With Test~1, we perform experiments according to the standard source extraction scenario from interference and noise as present in various papers (e.g., \cite{Williamson2016,Williamson2017,Hershey2016,Chen2017,Isik2016,Wang2014, Yu2017,Luo2017,wang2019deep,8664086}). With Test~2, we investigate the influence of destructive degradations, which require a reconstruction of the original desired signal, on the performance. Finally, in Test~3, we investigate the influence of interfering and destructive degradations at the same time on the performance.  

\section{Performance Evaluation}
\label{sec:peval}
\begin{figure}[!t]
\input{spectraplot.tex}
\vspace{-1em}
\caption{Excerpt of log-magnitude STFT spectra of desired speech, degraded by zeroing every fifth time-frame and frequency, and after processing with $\text{DF}_{3\times 3}$. The degradation in this figure was performed for illustration purposes only unlike the random time-frame zeroing in the data sets. }
\label{Fig:reconstruction}
\vspace{-1.4em}
\end{figure}
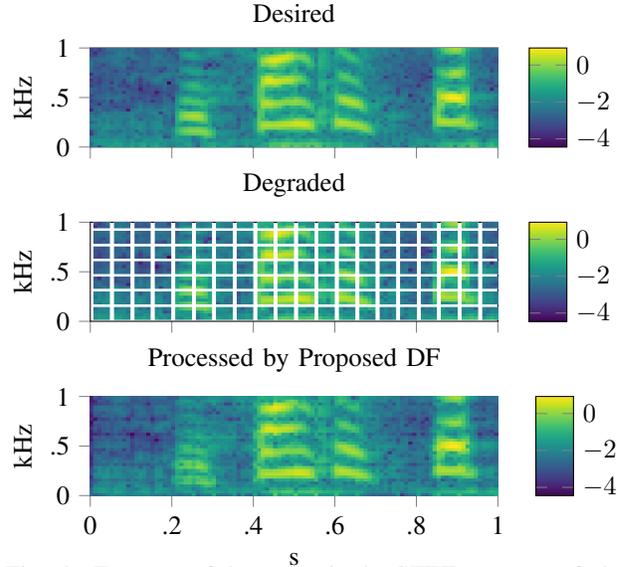

For performance evaluation\footnote{Audio examples are available at\\ www.audiolabs-erlangen.de/resources/2019-SPL-Deep-Filtering}, we used the signal-to-distortion-ratio (SDR), the signal-to-artifacts-ratio (SAR), the signal-to-interference-ratio (SIR) \cite{stoter2018,leroux2019}, the reconstruction MSE (see (\ref{Equ:comploss})), the short-time objective intelligibility (STOI) \cite{Taal2011,pystoi}, and the test data set. 
\begin{table}[!t]
\centering
\caption{Average results SDR, SIR, SAR, MSE (in dB), STOI for RM, CRM, and DF for test samples degraded with AudioSet interference in Test~1, with a notch-filter and time-frame zeroing (T-kill) in Test~2, and the combination in Test~3. The mask indices lin. and tanh specify the employed output activation for the respective model whereas the DF indices specify the employed filter dimension $(2\cdot L+1 \times 2\cdot I+1)$.} 
\label{tb:AvResults}
\begin{tabular}{lcccccc}
\toprule
\multicolumn{6}{l}{Test 1: Interference}\\
\midrule
 &						 SDR						 & SAR	 				 & SIR     				  & MSE  		& STOI     \\
 Input &  7.9 &  8.0 & \textbf{27.5} &   1.7 & 0.81 \\
 Proposed $\text{DF}_{5 \times 3}$ &15.4 & 20.6& 25.2 & -10.8 &\textbf{0.86} \\
 Proposed $\text{DF}_{3 \times 3} $&15.4 & \textbf{20.7} & 25.2& -10.8 & \textbf{0.86} \\
 Proposed $\text{DF}_{3 \times 5} $&15.4 & 20.5 & 25.4 & -10.8 & \textbf{0.86} \\
 Proposed $\text{DF}_{5 \times 1}$&\textbf{15.5} & \textbf{20.7} & 25.2 & -\textbf{11.1} & \textbf{0.86} \\
 Proposed $\text{DF}_{1 \times 5}$ & 15.3 & 20.2 & 25.4 & -10.8 & \textbf{0.86} \\
 $\text{CRM}_\text{tanh}$ & 15.1 & 19.8 & 24.7 & -10.7 & \textbf{0.86} \\
$\text{CRM}_\text{lin.}$ &  14.8 & 19.4 & 24.2 & -10.3 & 0.85 \\
 $\text{RM}_\text{tanh}$ &14.6& 18.0 & 25.6 & -10.2 & \textbf{0.86} \\
 $\text{RM}_\text{lin.}$&14.6& 17.8 & 25.4 & -10.1 & \textbf{0.86} \\
\\
  \multicolumn{6}{l}{Test 2: T-kill and Notch }\\
\midrule
 &						 SDR						 & SAR	 				 & SIR     				  & MSE  		& STOI     \\
Input& 11.5 & 12.3& 16.6 &  -7.8 & 0.89 \\
  Proposed $\text{DF}_{5 \times 3} $&22.8 & \textbf{34.5} & 25.6 & -\textbf{18.8} & \textbf{ 0.94} \\
  Proposed $\text{DF}_{3 \times 3} $&22.5 & 33.2 & 25.6 & -18.2 & \textbf{ 0.94} \\
 Proposed $\text{DF}_{3 \times 5} $& \textbf{23.0} & 34.1 & \textbf{26.0} & -18.7 &\textbf{ 0.94} \\
  Proposed $\text{DF}_{5\times 1}$&19.8 & 28.5 & 23.1 & -15.5& 0.93 \\
 Proposed $\text{DF}_{1 \times 5} $& 11.6 & 12.3 & 16.9 &  -7.9 & 0.89 \\
  $\text{CRM}_\text{tanh}$ &11.5 & 12.3 & 16.6 &  -7.8 & 0.89 \\
 $\text{CRM}_\text{lin.}$ & 11.5 & 12.3 & 16.6&  -7.8 & 0.89 \\
  $\text{RM}_\text{tanh}$ &11.5 & 12.3 & 16.6 &  -7.8 & 0.89 \\
  $\text{RM}_\text{lin.}$&11.5 & 12.3 & 16.7 &  -7.8 & 0.89 \\

\\
 \multicolumn{6}{l}{Test 3: Interference, T-kill, and Notch}\\ 
\midrule
 &						 SDR						 & SAR	 				 & SIR     				  & MSE  		& STOI     \\
 Input & 5.9 &  5.6 & 15.8&   1.0 & 0.76 \\
 Proposed $\text{DF}_{5 \times 3} $& \textbf{14.0} & \textbf{21.1} & 20.7 &  \textbf{-10.0}& \textbf{0.85} \\
  Proposed $\text{DF}_{3 \times 3} $&\textbf{14.0} & 20.8 & 20.7 & \textbf{-10.0} & \textbf{0.85} \\
  Proposed $\text{DF}_{3\times 5} $&\textbf{14.0} & 20.8 & \textbf{20.9} &  -9.9 & \textbf{0.85} \\
  Proposed $\text{DF}_{5 \times 1}$&13.5 & 20.2 & 19.3 &  -9.6 &\textbf{0.85} \\
   Proposed $\text{DF}_{1 \times 5}$ &9.6 & 12.3 & 15.1 &  -6.2 & 0.82 \\
   $\text{CRM}_\text{tanh}$ &9.5 & 12.2 & 14.7 &  -6.1 & 0.82 \\
  $\text{CRM}_\text{lin.}$ & 9.4 & 12.0 & 14.6 &  -6.0 & 0.81 \\
   $\text{RM}_\text{tanh}$ &9.4 & 11.5 & 14.8 &  -6.0 & 0.81 \\
   $\text{RM}_\text{lin.}$&9.4 & 11.5 & 14.8 &  -6.0 & 0.81 \\

\bottomrule
\end{tabular}
\vspace{-2em}
\end{table}
In Test~0, we evaluated the introduced distortion when clean speech is processed as many enhancement algorithms degrade the desired signal when the input SNR is very high. The resulting SDR was above 32~dB in SDR for all models. Hence, all DNNs can be applied even to clean speech as the introduced speech distortion is limited. Table \ref{tb:AvResults} shows the average results of Test~1~-~3. In Test~1, DFs, CRMs, and RMs showed to generalize well to unseen interference. Overall, the maximum SDR values and minimum MSE values were obtained using DFs. The best performing mask in terms of SDR is the CRM with tanh output activation with a distance between 0.2 and 0.4~dB of SDR to the DFs. In general, DFs yield a small improvement of SDR and MSE compared to mask-based enhancement in Test~1 as they are capable of reconstructing TF bins with destructive interference using bounded DNN output values. In Test~2, DFs outperformed CRMs and RMs as expected because the test conditions provided a comparable scenario to destructive interference. Processing with CRMs and RMs did not improve the signal quality as gains applied to zero (T-kill) cannot reconstruct the desired signals. The SDR did not decrease compared to the input similar to the results of Test~0. The only DF without improved results is $\text{DF}_{1\times 5}$ as its filters only incorporate several frequency bins of the same time-frame which have been zeroed by T-kill. Figure \ref{Fig:reconstruction} depicts log-magnitude spectra of clean speech, degraded speech by zeroing every fifth time-frame and frequency axis, and after enhancement with DF. Traces of the grid are still visible in low but not in high energy spectral regions as focused on by the loss in (\ref{Equ:comploss}). In Test~3, the DFs which considered temporal and spectral context performed best followed first by  $\text{DF}_{5\times 1}$ and then by $\text{DF}_{1\times 5}$ and eventually by the masks. When only temporal or spectral context is considered by DFs, the degradation introduced either by notch-filters ($\text{DF}_{5\times 1}$) or T-kill ($\text{DF}_{1\times 5}$) has a wider spread over the spectrum than the filters which violates (11) and consequently leads to worse reconstruction. Hence, the filter dimensions have to be chosen based on the spread of destructive degradations in the spectrum, so that (11) is not violated. Given no violation of (11), the DFs perform on par in Test~3.
\section{Conclusion}
We extended the concept of time-frequency masks for signal extraction to complex filters to increase the interference reduction and decrease the signal distortion and to address destructive interference of desired and undesired signals. We proposed to estimate the filters with a deep neural network trained end-to-end, which avoids defining ground-truth filters for training, which would be crucial due to the necessity to define filters for network training given infinity many possibilities consistently. All methods under test were able to perform speech extraction given unknown interference signals from AudioSet and did not degrade clean speech severely. The proposed deep filtering outperformed the mask baselines in all tests in terms of SDR, especially when packet-loss was simulated by time-frame zeroing and notch-filters were applied. Hence, with deep filters, signal extraction and/or reconstruction seems to be feasible under very adverse conditions such as packet-loss.
 \def\baselinestretch{1}
\balance
\bibliographystyle{IEEEtran}

\bibliography{sapref}



\end{document}

%% file: FiltervsMask.pdf_tex
\begingroup%
  \makeatletter%
  \providecommand\color[2][]{%
    \errmessage{(Inkscape) Color is used for the text in Inkscape, but the package 'color.sty' is not loaded}%
    \renewcommand\color[2][]{}%
  }%
  \providecommand\transparent[1]{%
    \errmessage{(Inkscape) Transparency is used (non-zero) for the text in Inkscape, but the package 'transparent.sty' is not loaded}%
    \renewcommand\transparent[1]{}%
  }%
  \providecommand\rotatebox[2]{#2}%
  \ifx\svgwidth\undefined%
    \setlength{\unitlength}{337.5bp}%
    \ifx\svgscale\undefined%
      \relax%
    \else%
      \setlength{\unitlength}{\unitlength * \real{\svgscale}}%
    \fi%
  \else%
    \setlength{\unitlength}{\svgwidth}%
  \fi%
  \global\let\svgwidth\undefined%
  \global\let\svgscale\undefined%
  \makeatother%
  \begin{picture}(1,0.77777778)%
    \put(0,0){\includegraphics[width=\unitlength,page=1]{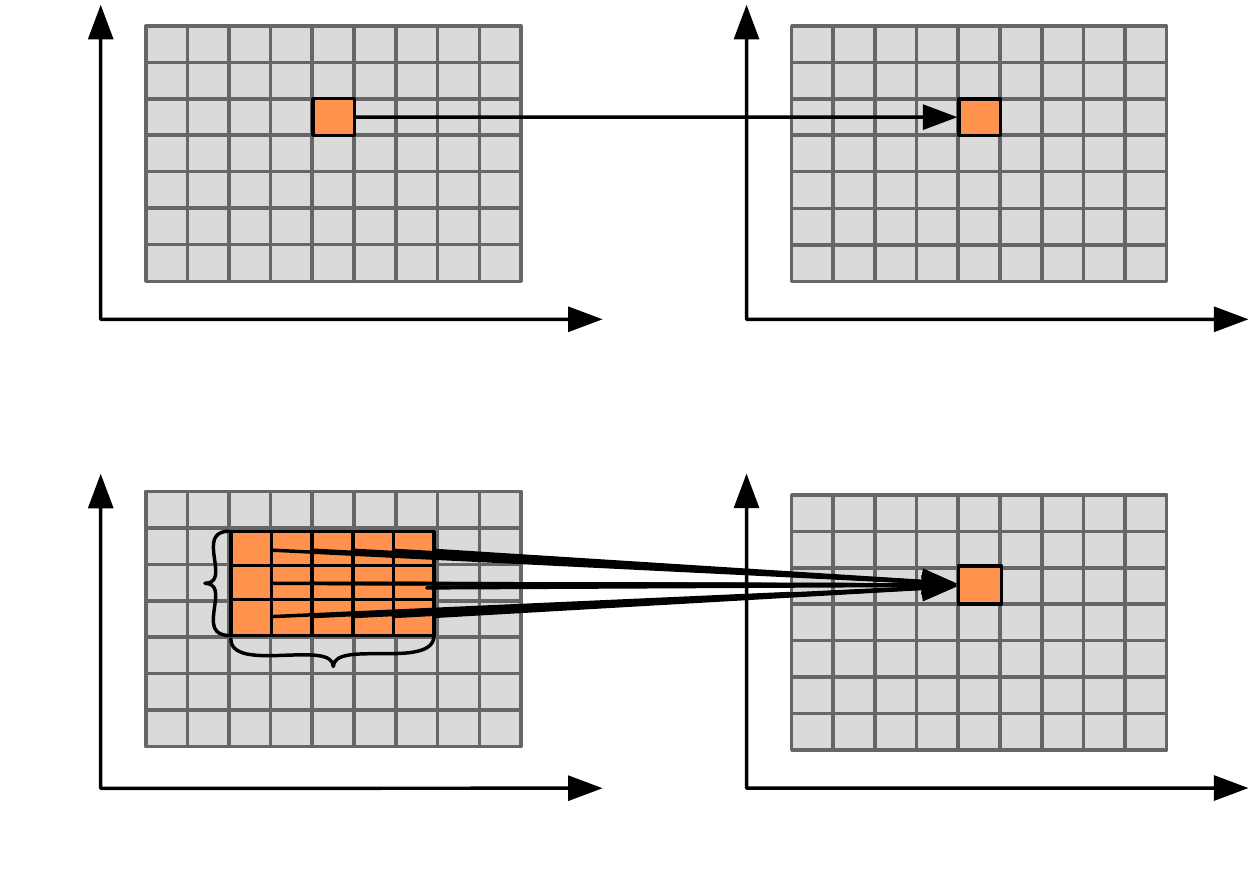}}%
    \put(0.46,0.28){\color[rgb]{0,0,0}\makebox(0,0)[lb]{\smash{Filter}}}%
    \put(0.22824464,0.72958726){\color[rgb]{0,0,0}\makebox(0,0)[lb]{\smash{Input}}}%
    \put(0.46,0.63){\color[rgb]{0,0,0}\makebox(0,0)[lb]{\smash{Mask}}}%
    \put(0.05,0.49){\color[rgb]{0,0,0}\rotatebox{90}{\makebox(0,0)[lb]{\smash{Freqcuency}}}}%
    \put(0.2272416,0.03613083){\color[rgb]{0,0,0}\makebox(0,0)[lb]{\smash{Time}}}%
    \put(0.05,0.13){\color[rgb]{0,0,0}\rotatebox{90}{\makebox(0,0)[lb]{\smash{Frequency}}}}%
    \put(0.74,0.72958726){\color[rgb]{0,0,0}\makebox(0,0)[lb]{\smash{Output}}}%
    \put(0.02367914,0.41){\color[rgb]{0,0,0}\makebox(0,0)[lb]{\smash{a)}}}%
    \put(0.02198197,0.03613083){\color[rgb]{0,0,0}\makebox(0,0)[lb]{\smash{b)}}}%
    \put(0.74,0.03613083){\color[rgb]{0,0,0}\makebox(0,0)[lb]{\smash{Time}}}%
    \put(0.74,0.41){\color[rgb]{0,0,0}\makebox(0,0)[lb]{\smash{Time}}}%
    \put(0.2272416,0.41){\color[rgb]{0,0,0}\makebox(0,0)[lb]{\smash{Time}}}%
    \put(0.2272416,0.34917786){\color[rgb]{0,0,0}\makebox(0,0)[lb]{\smash{Input}}}%
    \put(0.74,0.34917786){\color[rgb]{0,0,0}\makebox(0,0)[lb]{\smash{Output}}}%
    \put(0.145,0.203){\color[rgb]{0,0,0}\rotatebox{90}{\makebox(0,0)[lb]{\smash{${\scriptstyle 2 \cdot I +1}$}}}}%
    \put(0.5,0.5){\color[rgb]{0,0,0}\makebox(0,0)[lt]{\begin{minipage}{0.35\unitlength}\raggedright \end{minipage}}}%
    \put(0.22,0.15){\color[rgb]{0,0,0}\rotatebox{-0.15767191}{\makebox(0,0)[lb]{\smash{${\scriptstyle 2 \cdot L +1}$}}}}%
  \end{picture}%
\endgroup%

%% file: spectraplot.tex
\begin{tikzpicture}
\begin{groupplot}[group style={group size=1 by 3},  colorbar,colormap/viridis,
colorbar style={ylabel={}},
colormap/viridis,point meta max=0.926345705986023,
point meta min=-4.44641923904419]
\nextgroupplot[
height=2.9cm,
tick align=outside,
tick pos=left,
title={Desired},
width=7cm,
x grid style={white!69.01960784313725!black},
xmin=0, xmax=100,
y grid style={white!69.01960784313725!black},
ylabel={kHz},
ymin=0, ymax=1,xtick={0,20,40,60,80,100},
xticklabels={,,,,,},
ytick={0,0.5,1},
yticklabels={0,.5,1}
]
]
\addplot graphics [includegraphics cmd=\pgfimage,xmin=0, xmax=100, ymin=0, ymax=1] {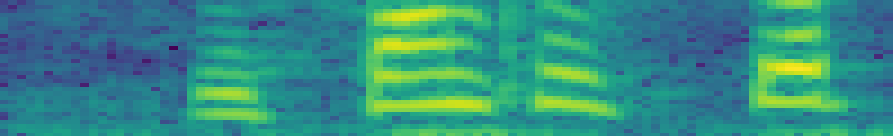};
\nextgroupplot[
height=2.9cm,
tick align=outside,
tick pos=left,
title={Degraded},
width=7cm,
x grid style={white!69.01960784313725!black},
xmin=0, xmax=100,
y grid style={white!69.01960784313725!black},
ymin=0, ymax=1,xtick={0,20,40,60,80,100},
ytick={0,0.5,1},
yticklabels={0,.5,1},
xticklabels={,,,,,},ylabel={kHz}
]
]
\addplot graphics [includegraphics cmd=\pgfimage,xmin=0, xmax=100, ymin=0, ymax=1] {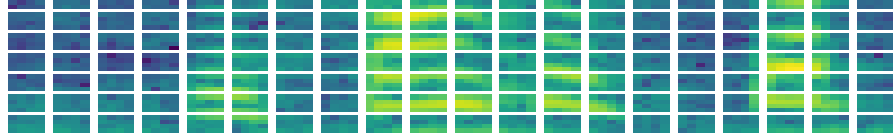};
\nextgroupplot[
height=2.9cm,
tick align=outside,
tick pos=left,
title={Processed by Proposed DF },
point meta max=0.926345705986023,
point meta min=-4.44641923904419,
width=7cm,
x grid style={white!69.01960784313725!black},
xlabel={s},
xmin=0, xmax=100,
y grid style={white!69.01960784313725!black},
ylabel={kHz},
ymin=0, ymax=1,
xtick={0,20,40,60,80,100},
xticklabels={0,.2,.4,.6,.8,1},
ytick={0,0.5,1},
yticklabels={0,.5,1},
,ylabel={kHz}]
\addplot graphics [includegraphics cmd=\pgfimage,xmin=0, xmax=100, ymin=0, ymax=1] {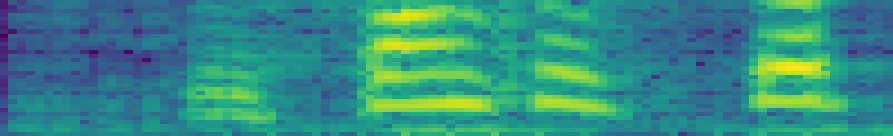};
\end{groupplot}

\end{tikzpicture}

%% file: DFARXIVV3.bbl
\begin{thebibliography}{10}
\providecommand{\url}[1]{#1}
\csname url@samestyle\endcsname
\providecommand{\newblock}{\relax}
\providecommand{\bibinfo}[2]{#2}
\providecommand{\BIBentrySTDinterwordspacing}{\spaceskip=0pt\relax}
\providecommand{\BIBentryALTinterwordstretchfactor}{4}
\providecommand{\BIBentryALTinterwordspacing}{\spaceskip=\fontdimen2\font plus
\BIBentryALTinterwordstretchfactor\fontdimen3\font minus
  \fontdimen4\font\relax}
\providecommand{\BIBforeignlanguage}[2]{{%
\expandafter\ifx\csname l@#1\endcsname\relax
\typeout{** WARNING: IEEEtran.bst: No hyphenation pattern has been}%
\typeout{** loaded for the language `#1'. Using the pattern for}%
\typeout{** the default language instead.}%
\else
\language=\csname l@#1\endcsname
\fi
#2}}
\providecommand{\BIBdecl}{\relax}
\BIBdecl

\bibitem{Tan2019}
K.~Tan and D.~Wang, ``Complex spectral mapping with a convolutional recurrent
  network for monaural speech enhancement,'' in \emph{Proc. {IEEE} Intl. Conf.
  on Acoustics, Speech and Signal Processing (ICASSP)}, May 2019, pp.
  6865--6869.

\bibitem{Han2015}
K.~Han, Y.~Wang, D.~Wang, W.~S. Woods, I.~Merks, and T.~Zhang, ``Learning
  spectral mapping for speech dereverberation and denoising,'' \emph{{IEEE/ACM}
  Trans. Audio, Speech, Lang. Process.}, vol.~23, no.~6, pp. 982--992, Jun.
  2015.

\bibitem{Wang2014}
Y.~Wang, A.~Narayanan, and D.~Wang, ``On training targets for supervised speech
  separation,'' \emph{{IEEE/ACM} Trans. Audio, Speech, Lang. Process.},
  vol.~22, no.~12, pp. 1849--1858, Dec. 2014.

\bibitem{Williamson2016}
D.~S. Williamson, Y.~Wang, and D.~Wang, ``Complex ratio masking for monaural
  speech separation,'' \emph{{IEEE} Trans. Audio, Speech, Lang. Process.},
  vol.~24, no.~3, pp. 483--492, Aug. 2016.

\bibitem{Williamson2017}
D.~S. Williamson and D.~Wang, ``Speech dereverberation and denoising using
  complex ratio masks,'' in \emph{Proc. {IEEE} Intl. Conf. on Acoustics, Speech
  and Signal Processing (ICASSP)}, Mar. 2017, pp. 5590--5594.

\bibitem{Hershey2016}
J.~R. Hershey, Z.~Chen, J.~L. Roux, and S.~Watanabe, ``Deep clustering:
  Discriminative embeddings for segmentation and separation,'' in \emph{Proc.
  {IEEE} Intl. Conf. on Acoustics, Speech and Signal Processing (ICASSP)}, Mar.
  2016, pp. 31--35.

\bibitem{Chen2017}
Z.~Chen, Y.~Luo, and N.~Mesgarani, ``Deep attractor network for
  single-microphone speaker separation,'' in \emph{Proc. {IEEE} Intl. Conf. on
  Acoustics, Speech and Signal Processing (ICASSP)}, Mar. 2017, pp. 246--250.

\bibitem{Isik2016}
Y.~Isik, J.~L. Roux, Z.~Chen, S.~Watanabe, and J.~R. Hershey, ``Single-channel
  multi-speaker separation using deep clustering,'' in \emph{Proc. Interspeech
  Conf.}, Sep. 2016, pp. 545--549.

\bibitem{Yu2017}
D.~Yu, M.~Kolbæk, Z.~H. Tan, and J.~Jensen, ``Permutation invariant training
  of deep models for speaker-independent multi-talker speech separation,'' in
  \emph{Proc. {IEEE} Intl. Conf. on Acoustics, Speech and Signal Processing
  (ICASSP)}, Mar. 2017, pp. 241--245.

\bibitem{Luo2017}
Y.~Luo, Z.~Chen, J.~R. Hershey, J.~L. Roux, and N.~Mesgarani, ``Deep clustering
  and conventional networks for music separation: {Stronger} together,'' in
  \emph{Proc. {IEEE} Intl. Conf. on Acoustics, Speech and Signal Processing
  (ICASSP)}, Mar. 2017, pp. 61--65.

\bibitem{wang2019deep}
Z.-Q. Wang, K.~Tan, and D.~Wang, ``Deep learning based phase reconstruction for
  speaker separation: {A} trigonometric perspective,'' in \emph{Proc. {IEEE}
  Intl. Conf. on Acoustics, Speech and Signal Processing (ICASSP)}, May 2019,
  pp. 71--75.

\bibitem{8664086}
J.~{Le Roux}, G.~{Wichern}, S.~{Watanabe}, A.~{Sarroff}, and J.~R. {Hershey},
  ``Phasebook and friends: {Leveraging} discrete representations for source
  separation,'' \emph{{IEEE} J. sel. Top. in Sig. Proc.}, vol.~13, no.~2, pp.
  370--382, May 2019.

\bibitem{Kolbaek2017}
M.~Kolbaek, D.~Yu, Z.-H. Tan, J.~Jensen, M.~Kolbaek, D.~Yu, Z.-H. Tan, and
  J.~Jensen, ``Multitalker speech separation with utterance-level permutation
  invariant training of deep recurrent neural networks,'' \emph{{IEEE} Trans.
  Audio, Speech, Lang. Process.}, vol.~25, no.~10, pp. 1901--1913, Oct. 2017.

\bibitem{Mack2018}
W.~Mack, S.~Chakrabarty, F.-R. St{\"o}ter, S.~Braun, B.~Edler, and E.~A.~P.
  Habets, ``Single-channel dereverberation using direct {MMSE }optimization and
  bidirectional {LSTM} networks,'' in \emph{Proc. Interspeech Conf.}, Sep.
  2018, pp. 1314--1318.

\bibitem{Erdogan2018}
H.~Erdogan and T.~Yoshioka, ``Investigations on data augmentation and loss
  functions for deep learning based speech-background separation,'' in
  \emph{Proc. Interspeech Conf.}, Sep. 2018, pp. 3499--3503.

\bibitem{Wang2005a}
D.~Wang, ``On ideal binary mask as the computational goal of auditory scene
  analysis,'' in \emph{Speech Separation by Humans and Machines}, P.~Divenyi,
  Ed.\hskip 1em plus 0.5em minus 0.4em\relax Kluwer Academic, 2005, pp.
  181--197.

\bibitem{Hummersone20142}
C.~Hummersone, T.~Stokes, and T.~Brookes, ``On the ideal ratio mask as the goal
  of computational auditory scene analysis,'' in \emph{Blind Source
  Separation}, G.~R. Naik and W.~Wang, Eds.\hskip 1em plus 0.5em minus
  0.4em\relax Springer, 2014, pp. 349--368.

\bibitem{Mayer2017}
F.~Mayer, D.~S. Williamson, P.~Mowlaee, and D.~Wang, ``Impact of phase
  estimation on single-channel speech separation based on time-frequency
  masking,'' \emph{J. Acoust. Soc. Am.}, vol. 141, no.~6, pp. 4668--4679, 2017.

\bibitem{Weninger2015}
F.~Weninger, H.~Erdogan, S.~Watanabe, E.~Vincent, J.~Roux, J.~R. Hershey, and
  B.~Schuller, ``Speech enhancement with {LSTM} recurrent neural networks and
  its application to noise-robust {ASR},'' in \emph{Proc. of the 12th Int.
  Conf. on Lat. Var. An. and Sig. Sep.}, ser. LVA/ICA.\hskip 1em plus 0.5em
  minus 0.4em\relax New York, USA: Springer-Verlag, 2015, pp. 91--99.

\bibitem{Xu2017}
X.~Li, J.~Li, and Y.~Yan, ``Ideal ratio mask estimation using deep neural
  networks for monaural speech segregation in noisy reverberant conditions,''
  in \emph{Proc. Interspeech Conf.}, Aug. 2017, pp. 1203--1207.

\bibitem{Benesty2011}
J.~Benesty, J.~Chen, and E.~A.~P. Habets, \emph{Speech Enhancement in the
  {STFT} Domain}, ser. SpringerBriefs in Electrical and Computer
  Engineering.\hskip 1em plus 0.5em minus 0.4em\relax Spring{\-}er-Ver{\-}lag,
  2011.

\bibitem{Benesty2011a}
J.~Benesty and Y.~Huang, ``A single-channel noise reduction {MVDR} filter,'' in
  \emph{Proc. {IEEE} Intl. Conf. on Acoustics, Speech and Signal Processing
  (ICASSP)}, May 2011, pp. 273--276.

\bibitem{Fischer2016}
D.~Fischer, S.~Doclo, E.~A.~P. Habets, and T.~Gerkmann, ``Combined
  single-microphone {Wiener} and {MVDR} filtering based on speech interframe
  correlations and speech presence probability,'' in \emph{Speech
  Communication; 12. ITG Symposium}, Oct. 2016, pp. 1--5.

\bibitem{Doerte2018}
D.~Fischer and S.~Doclo, ``Robust constrained {MFMVDR} filtering for
  single-microphone speech enhancement,'' in \emph{Proc. Intl. Workshop Acoust.
  Signal Enhancement ({IWAENC})}, Sep. 2018, pp. 41--45.

\bibitem{Hochreiter1997}
S.~Hochreiter and J.~Schmidhuber, ``Long short-term memory,'' \emph{Neural
  Computation}, vol.~9, no.~8, pp. 1735--1780, Nov. 1997.

\bibitem{Kingma2015}
J.~B. D.~Kingma, ``Adam: {A} method for stochastic optimization,'' in
  \emph{Proc. {IEEE} Intl. Conf. on Learn. Repr. (ICLR)}, May 2015, pp. 1--15.

\bibitem{Srivastava2014}
N.~Srivastava, G.~Hinton, A.~Krizhevsky, I.~Sutskever, and R.~Salakhutdinov,
  ``Dropout: A simple way to prevent neural networks from overfitting,''
  \emph{J. Mach. Learn. Res.}, vol.~15, no.~1, pp. 1929--1958, Jan. 2014.

\bibitem{gemmeke2017}
J.~F. Gemmeke, D.~P.~W. Ellis, D.~Freedman, A.~Jansen, W.~Lawrence, R.~C.
  Moore, M.~Plakal, and M.~Ritter, ``Audio {Set}: {An} ontology and
  human-labeled dataset for audio events,'' in \emph{Proc. {IEEE} Intl. Conf.
  on Acoustics, Speech and Signal Processing (ICASSP)}, Mar. 2017, pp.
  776--780.

\bibitem{Libri2015}
V.~Panayotov, G.~Chen, D.~Povey, and S.~Khudanpur, ``Librispeech: An {ASR}
  corpus based on public domain audio books,'' in \emph{Proc. {IEEE} Intl.
  Conf. on Acoustics, Speech and Signal Processing (ICASSP)}, Apr. 2015, pp.
  5206--5210.

\bibitem{stoter2018}
F.-R. St{\"o}ter, A.~Liutkus, and N.~Ito, ``The 2018 signal separation
  evaluation campaign,'' in \emph{Proc. of the 14th Int. Conf. on Lat. Var. An.
  and Sig. Sep. {(LVA/ICA)}}, Apr. 2018, pp. 293--305.

\bibitem{leroux2019}
J.~Le~Roux, S.~Wisdom, H.~Erdogan, and J.~R. Hershey, ``{SDR}--half-baked or
  well done?'' in \emph{Proc. {IEEE} Intl. Conf. on Acoustics, Speech and
  Signal Processing (ICASSP)}, May 2019, pp. 626--630.

\bibitem{Taal2011}
C.~H. Taal, R.~C. Hendriks, R.~Heusdens, and J.~Jensen, ``An algorithm for
  intelligibility prediction of time-frequency weighted noisy speech,''
  \emph{{IEEE} Trans. Audio, Speech, Lang. Process.}, vol.~19, no.~7, pp.
  2125--2136, Sep. 2011.

\bibitem{pystoi}
M.~Pariente, ``pystoi,'' https://github.com/mpariente/pystoi, 2018.

\end{thebibliography}
